\documentclass[traditabstract]{aa}

\usepackage{graphicx}
\usepackage{txfonts}
\usepackage{float}

\usepackage{natbib}
\bibpunct{(}{)}{;}{a}{}{,}

\usepackage{hyperref}
\hypersetup{colorlinks=true,
            citecolor=blue,
            linkcolor=blue}

\def\vv#1{\vec{#1}}
\def \be {\begin{equation}}
\def \ee {\end{equation}}
\def\om{\omega}
\def\vk{\vv{\rm k}}
\def\vs{\vv{\rm s}}
\def\vr{\vv{r}}

\def\vL{\vv{L}}
\def\vG{\vv{G}}
\def\m{m_{01}}
\def\mj{M_\mathrm{Jup}}
\def\C{C}
\def\G{{\cal G}}
\def\H{H}
\def\K{K}
\def\vK{\vv{\K}}
\def\ui{u}
\def\vi{v}
\def\uo{u_*}
\def\vo{v_*}
\def\uic{\ui_c}
\def\uoc{\uo^c}
\def\x{x}
\def\y{y}
\def\z{z}
\def\zx{Z(\x,\ui)}
\def\zxs{Z^2(\x, \ui)}
\def\gx{G(\x,\ui)}
\def\zy{Z_*(\y,\uo)}
\def\zys{Z_*^2(\y,\uo)}
\def\gy{G_*(\y, \uo)}

\def\figpath{}
\def \llabel#1{\label{#1}}

\usepackage{color}

\begin{document}

\title{Stellar and planetary Cassini states}
\titlerunning{Stellar and planetary Cassini states}

\author{
Alexandre C. M. Correia\inst{1,2}
}

 
\institute{CIDMA, Departamento de F\'isica, Universidade de Aveiro, Campus de
Santiago, 3810-193 Aveiro, Portugal
  \and 
ASD, IMCCE-CNRS UMR8028,
Observatoire de Paris,
77 Av. Denfert-Rochereau, 75014 Paris, France  
}

\date{Received ; accepted To be inserted later}

\abstract{Cassini states correspond to equilibria of the spin axis of a body when its orbit is perturbed. 
They were initially described for satellites, but the spin axis of stars and planets undergoing strong dissipation can also evolve into some equilibria.
For small satellites, the rotational angular momentum is usually much smaller than the total angular momentum, so classical methods for finding Cassini states rely on this approximation.
Here we present a more general approach, which is valid for the secular quadrupolar non-restricted problem with spin.
Our method is still valid when the precession rate and the mutual inclination of the orbits are not constant. 
Therefore, it can be used to study stars with close-in companions, or planets with heavy satellites, like the Earth-Moon system.
}


   \maketitle
%



\section{Introduction}

Following observations of the Moon, \citet{Cassini_1693} established three empirical laws on its rotational motion.
The first stated that the rotation rate and the orbital mean motion are synchronous, the second that the angle between Moon's equator and the ecliptic is constant, and the third that the Moon's spin axis and the normals to its orbital plane and ecliptic remain coplanar.
The observed physical librations are described as departures of the rotational motion from these three equilibrium laws.

\citet{Colombo_1966} has shown that the second and third laws are independent of the first one, in the sense that even if the rotation rate is not synchronous, the second and third laws can still be satisfied since they correspond to the minimum dissipation of energy for the spin axis. 
For a non-synchronous Moon, only the angle between its equator and the ecliptic would change.
Indeed, while the first law requires an triaxial ellipsoid to work, the
two other laws only require an oblate spheroid.
Moreover, \citet{Colombo_1966} generalised his theory to any satellite or planet whose nodal line on the invariant plane shifts because of perturbations, which can have a different origin, such as the oblateness of the central body, perturbations from a third body, or both.
\citet{Peale_1969} further generalised second and third laws to include the effects of an axial asymmetry and rotation rates commensurable with the orbital mean motion.

In the classical approach, the Hamiltonian of a slightly aspherical body is developed in a reference frame that precesses with the orbit.
If the angular momentum and the energy are approximately conserved, the precession of the spin axis relative to the coordinate system fixed in the orbital plane is determined by the intersection of a sphere and a parabolic cylinder.
The spin axis is fixed relative to the precessing orbit when the energy has an extreme value.
Thus, these equilibria states for the spin axis can be the end point of dissipation, usually with a tidal origin \citep{Ward_1975}, and they received the name of Cassini states.

An important hypothesis required to apply the classical approximation is that the orbit precesses about an inertial plane with uniform angular velocity and constant inclination.
For small satellites, this requirement is usually met; they keep a constant inclination either to the ecliptic (like the Moon) or to the equatorial bulge of the central planet.
Two planets in the solar system are also expected to occupy Cassini states, namely Mercury \citep{Peale_2006} and Venus \citep{Correia_Laskar_2003I}. 
The classical hypothesis still works on these cases, because none of these planets have satellites.
However, for a planet with a huge satellite (such as the Earth-Moon or the Pluto-Charon systems), the classical approximations fail for several orbital configurations \citep{Boue_Laskar_2006}.
It also fails if one wants to inspect the spin of a star with high-mass close-in planetary companions.

In this paper we intend to generalise the theory of Cassini states further. 
For simplicity we consider the secular three-body hierarchical problem with spin, since it has been shown to be integrable \citep{Touma_Wisdom_1994, Boue_Laskar_2006}. 
It thus allows us to compute the level curves of the Hamiltonian.
In section~2 we revise the secular dynamics of the three-body problem and introduce a new set of variables that allows computing Cassini states straightforwardly.
In section~3 we present the general conditions to obtain the Cassini states and apply it to the case of a star with two planetary companions and to the Earth-Moon system.
In section~4 we derive some conclusions and explain how to extend the model to n-body systems.


\section{Secular dynamics}

\begin{figure}
\begin{center}
\includegraphics[width=\columnwidth]{\figpath 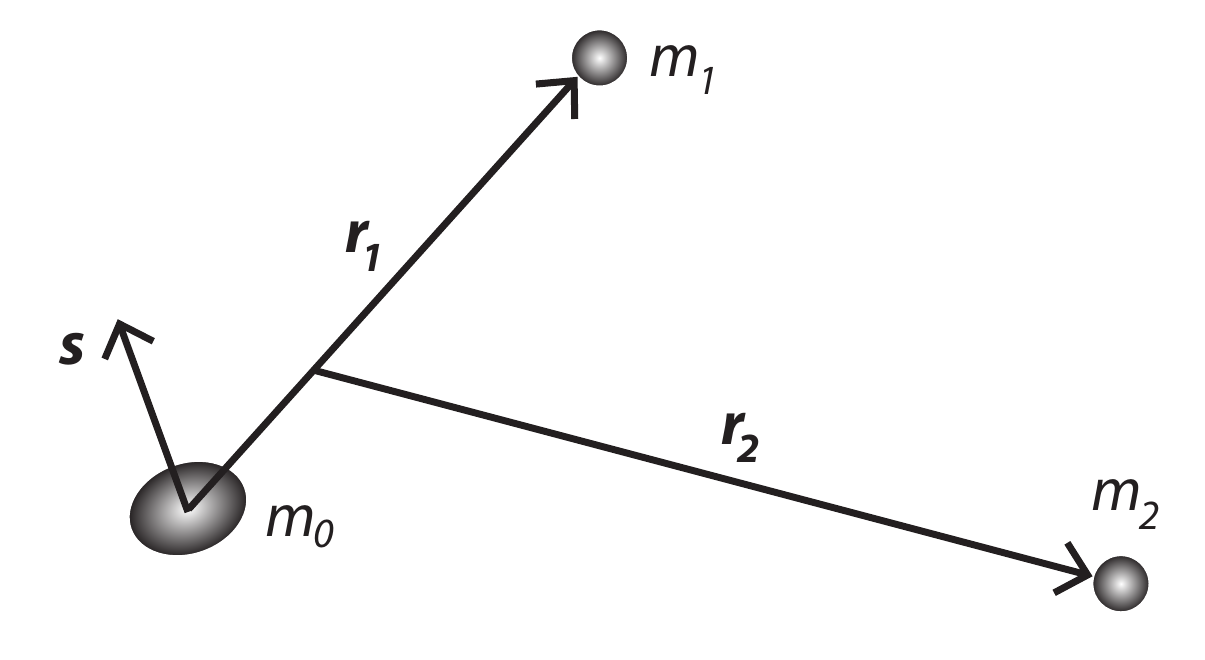} 
 \caption{Jacobi coordinates, where $\vr_1 $ is the position of $m_1$ relative to $m_0$ (inner orbit), and $ \vr_2 $ the position of $m_2$ relative to the center of mass of $m_0$ and $m_1$ (outer orbit). The body with mass $m_0$ is considered an oblate ellipsoid, where $\vs$ is the spin axis. \llabel{fig1}  }
\end{center}
\end{figure}

We consider a three-body hierarchical system composed of a central pair with masses $m_0$ and $m_1$, together with an external companion with mass $m_2$ (Fig.\,\ref{fig1}).
The body with mass $m_0$ is considered an oblate ellipsoid with gravity field coefficient $J_2$ and rotation rate $\vv{\om} = \om \, \vs $, where $\vs$ is also the axis of maximal inertia (gyroscopic approximation) with moment $\C$.
The rotational angular momentum is then simply given by 
\be
\vL = L \, \vs = \C \om \, \vs \ . \llabel{141216a}
\ee 
For the orbits we use Jacobi canonical coordinates, with $\vr_1 $  the position of $m_1$ relative to $m_0$ (inner orbit), and $ \vr_2 $ the position of $m_2$ relative to the centre of mass of $m_0$ and $m_1$ (outer orbit).
The Keplerian orbital angular momenta of each orbit are 
\be
\vG_i = G_i \, \vk_i = \beta_i \sqrt{\mu_i a_i (1-e_i^2)} \, \vk_i 
\ , \llabel{141216b}
\ee
where $\vk_i$ is the unit vector normal to the orbital plane ($i = 1, 2$), 
$a_i$ is the semi-major axis, $e_i$ is the eccentricity, 
$ \beta_1 = m_0 m_1 / \m $ and $ \beta_2 = \m m_2 / (\m +m_2)$ are the reduced masses, 
$\mu_1 = \G \m $,  $\mu_2 = \G (\m + m_2)$, with $\m = (m_0 + m_1) $, and $\G$ is the gravitational constant.

\subsection{Hamiltonian}

For simplicity, we restrict our analysis to hierarchical systems for which $r_1 \ll r_2$, where 
$r_i = ||\vr_i||$.
Similarly, we assume that the ellipsoid mean radius is $R \ll r_1$.
Thus, the interaction potential between the two orbits and the body oblateness can be restricted to terms in $(r_1/r_2)^2$ and  $(R/r_i)^2$; that is to say, we adopt a quadrupolar approximation for the Hamiltonian \citep[e.g.,][]{Touma_Wisdom_1994}.

Since we are only concerned with the secular evolution of the spin axis, we average the Hamiltonian of the quadrupolar three-body problem over the mean anomalies of the two orbits and also over the argument of perihelion of the inner orbit.
For the non-constant parts, we simply get \citep[][]{Goldreich_1966a, Touma_Wisdom_1994, Boue_Laskar_2006} 
\be
\H = 
-\frac{\alpha_1}{2} (\vs \cdot \vk_1)^2 
-\frac{\alpha_2}{2} (\vs \cdot \vk_2)^2 
-\frac{\gamma}{2} (\vk_1 \cdot \vk_2)^2
\ , \llabel{141216c}
\ee
where
\be 
\alpha_i =  \frac{3 \G m_0 m_i J_2 R^2}{2 a_i^3 (1-e_i^2)^{3/2}} \ ,
\quad \mathrm{and} \quad
\gamma = \frac{3 \G \beta_1 m_2 a_1^2 (2 + 3 e_1^2)}{8 a_2^3 (1-e_2^2)^{3/2}} \ .
\llabel{141216d}
\ee

\subsection{Equations of motion}

In the secular conservative problem, all quantities appearing in the Hamiltonian (Eq.\,(\ref{141216c})) are constant, except for the unit vectors $\vs$ and $\vk_i$, which can be related to the angular moment components (Eqs.\,(\ref{141216a}), (\ref{141216b})).
The evolution of the system can therefore be described by the evolution of the angular momentum components, which can be obtained from the Hamiltonian through Poisson brackets \citep[e.g.,][]{Dullin_2004, Breiter_etal_2005b}
\be
\frac{d \vL}{d t} = \{ \vL, \H \} = \nabla_{\vL} \, \H \times \vL \ ,
\quad
\frac{d \vG_i}{d t} = \{ \vG_i, \H \} = \nabla_{\vG_i} \, \H \times \vG_i 
\ , \llabel{141216e}
\ee
which gives for the unit vectors
\be
\frac{d \vs}{d t} = 
- \frac{\alpha_1}{L} (\vs \cdot \vk_1) \, \vk_1 \times \vs 
- \frac{\alpha_2}{L} (\vs \cdot \vk_2) \, \vk_2 \times \vs
\ , \llabel{141216g}
\ee
and
\be
\frac{d \vk_i}{d t} = 
-\frac{\alpha_i}{G_i} (\vs \cdot \vk_i) \, \vs \times \vk_i
-\frac{\gamma}{G_i} (\vk_i \cdot \vk_j) \, \vk_j \times \vk_i
\ , \llabel{141216h}
\ee
with $i \ne j = 1, 2 $.
From previous expressions we also see that the total angular momentum is conserved:
\be
\vK = \vL + \vG_1 + \vG_2 = const.
\llabel{141216i}
\ee

\subsection{Reduced problem}

\begin{figure}
\begin{center}
\includegraphics[width=\columnwidth]{\figpath 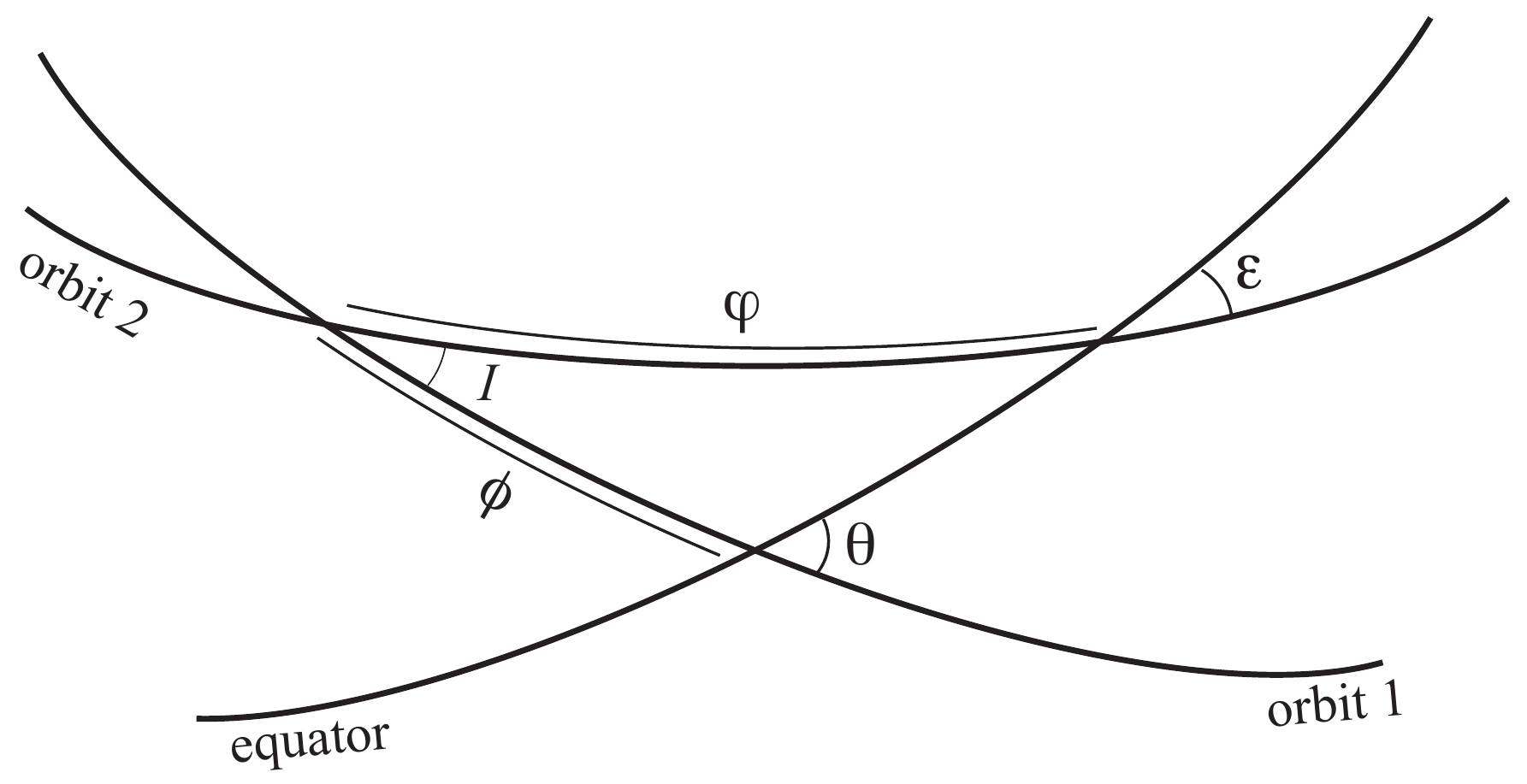} 
 \caption{Reference planes for the definition of the direction cosines and the precession angles.  \llabel{fig2}  }
\end{center}
\end{figure}

Following \citet{Goldreich_1966a} and \citet{Boue_Laskar_2006}, the equations of motion can be simplified if we consider only the relative position in space of the unit vectors $\vs$, $\vk_1$, and $\vk_2$, given by the direction cosines\footnote{The variables $x$ and $y$ here are switched with respect to those in \citet{Boue_Laskar_2006}, because they define the index 2 for the inner orbit and 1 for the outer orbit. They also differ from the notations in \citet{Goldreich_1966a}, where $x = \vs \cdot \vk_2$, $y = \vk_1 \cdot \vk_2$, and $z = \vs \cdot \vk_1$.} (Fig.\,\ref{fig2})
\be
x = \cos \theta = \vs \cdot \vk_1 \ , \quad 
y = \cos \varepsilon = \vs \cdot \vk_2 \ , \quad 
z = \cos I = \vk_1 \cdot \vk_2 \ ,
\llabel{141216j}
\ee
together with the ``berlingot'' shaped volume
\be
w = \vs \cdot (\vk_1 \times \vk_2) = \pm \sqrt{1- x^2 - y^2 - z^2 + 2xyz} 
\ . \llabel{141216k}
\ee
The equations of motion (\ref{141216g}) and (\ref{141216h}) are then rewritten as
\begin{eqnarray}
\dot x = \left(\frac{\gamma}{G_1} z - \frac{\alpha_2}{L} y \right) w \ , \quad
\dot y = \left(\frac{\alpha_1}{L} x -\frac{\gamma}{G_2} z \right) w \ , \nonumber
\end{eqnarray}
\be
\dot z =  \left(\frac{\alpha_2}{G_2} y - \frac{\alpha_1}{G_1} x  \right) w \ ,
\llabel{141216l}
\ee
and
\be
\dot w =  (yz-x) \frac{\dot x}{w} + (xz-y) \frac{\dot y}{w} + (xy-z) \frac{\dot z}{w} \ .
\llabel{141216m}
\ee
We can get $w$ directly from expression (\ref{141216k}), but this last equation can be useful for determining whether $w$ is positive or negative.
In addition, we still have two remaining integrals, one from the Hamiltonian (Eq.\,(\ref{141216c})),
\be
\H_0 = \alpha_1 x^2 + \alpha_2 y^2 + \gamma z^2 = - 2 \H 
\ , \llabel{141216n}
\ee
and another from the total angular momentum (Eq.\,(\ref{141216i})),
\be
\K_0 = L \, G_1 x + L \, G_2 y + G_1 G_2 z = (\K^2 - L^2 - G_1^2 - G_2^2)/2
\ , \llabel{141216o}
\ee
so equations (\ref{141216l}) reduce to an integrable problem. 

\subsection{New reduction}

\llabel{newvars}

The set of variables ($x,y,z$) allow us to find an integrable solution for the spin-orbit problem, but this solution is not straightforward \citep[see section 3.4 in][]{Boue_Laskar_2006}.
Therefore, we introduce here a new set of variables ($\ui,\vi$) that are more intuitive and natural to the Cassini states' problem.

\subsubsection{Projection on the inner orbit}

We let
\be
\ui = \frac{y - xz}{\sqrt{1-z^2}} = \sin \theta \cos \phi \ , 
\llabel{141217a}
\ee
and
\be
\vi = \frac{w}{\sqrt{1-z^2}} = \sin \theta \sin \phi \ ,
\llabel{141217b}
\ee
where $\phi$ is the angle measured along the inner orbit from the interception with the outer orbit to the interception with the equatorial plane (Fig.\,\ref{fig2}).
Thus, when $\phi = 0$ the unit vectors normal to these planes ($\vs$, $\vk_1$, $\vk_2$) lie in the same plane.
With this choice, $\x$ only depends on the new variables\footnote{We consider here that $\x>0$, but this method is still valid for $\x<0$ adopting $\x=-\sqrt{1 - \ui^2 - \vi^2}$.}
\be
x = \sqrt{1 - u^2 - v^2}  \ , \llabel{141217c}
\ee
while $\y$ and  $\z$ still depend on each other
\be
y = xz + \ui \sqrt{1 - z^2} \ . \llabel{141217d}
\ee
However, replacing $\y$ above in expression (\ref{141216o}), we get
\be
(G_1 + L \, x) \, z + L \, \ui \sqrt{1 - z^2} = (\K_0 - L \, G_1 x)/G_2 
\ , \llabel{141217e}
\ee
which can be explicitly solved for $\z$ as
\be
z = \frac{(G_1 + L \, x) \zx - L \, \ui \sqrt{1-\zxs}}{\gx}
\ , \llabel{141217f}
\ee
with
\be
\zx = \frac{\K_0 - L \, G_1 x}{G_2 \, \gx} 
\ , 
\llabel{141224a0}
\ee
\be
\gx = \sqrt{(G_1 + L \, x)^2+(L \, u)^2}
\ . \llabel{141224a}
\ee
Therefore, $\y$ and $\z$ also depend only on ($\x, \ui$), hence on the new variables ($\ui,\vi$), as well as the Hamiltonian  (Eq.\,(\ref{141216n}))
\be
\H_0 = \H_0 (\x, \ui,\K_0) = \H_0 (\ui,\vi,\K_0) \ . \llabel{141217g}
\ee
The corresponding equations of motion are
\begin{eqnarray}
\dot \ui &=& \left[ \frac{\alpha_1 \x + \alpha_2 \y \z}{L} - \frac{\gamma \z}{G_2}  - \frac{\gamma \z^2}{G_1}   + \frac{\y \z - \x}{1-\z^2} \left(\frac{\alpha_2}{G_2} \y - \frac{\alpha_1}{G_1} \x  \right) \right] \vi
\ , \llabel{141227d}
\end{eqnarray}
\begin{eqnarray}
\dot \vi &=& \left(\frac{\gamma}{G_1} z - \frac{\alpha_2}{L} y \right) \frac{yz-x}{\sqrt{1-z^2}} - \left(\frac{\alpha_1}{L} x -\frac{\gamma}{G_2} z \right) \ui \nonumber \\
&& + \left(\frac{\alpha_2}{G_2} y - \frac{\alpha_1}{G_1} x  \right) \frac{xy+ z \, (v^2-1)}{\sqrt{1-z^2}}
\ . \llabel{141227e}
\end{eqnarray}

\subsubsection{Projection on the outer orbit}

\llabel{projout}

A similar reduction ($\uo,\vo$) could be obtained for the projection of the spin on the outer orbit, by defining 
\be
\uo = \frac{yz-x}{\sqrt{1-z^2}} = \sin \varepsilon \cos \varphi \ ,
 \llabel{141230a}
\ee
\be
\vo = \vi = \frac{w}{\sqrt{1-z^2}}  = \sin \varepsilon \sin \varphi  \ ,
 \llabel{141230g}
\ee
where $\varphi$ is the angle measured along the outer orbit from the interception with the inner orbit to the interception with the equatorial plane (Fig.\,\ref{fig2}).
Thus, when $\varphi = 0$ the unit vectors normal to these planes 
still lie in the same plane.
With this choice, the Hamiltonian (\ref{141216n}) can also be expressed in the new variables
\be
\H_0 = \H_0 (\y, \uo,\K_0) = \H_0 (\uo,\vo,\K_0) \ , \llabel{141230b}
\ee
using the transformations
\be
y = \sqrt{1 - \uo^2 - \vo^2}  \ , \llabel{141230c}
\ee
\be
z = \frac{(G_2 + L \, y) \zy + L \, \uo \sqrt{1-\zys}}{\gy}
\ , \llabel{141230f}
\ee
\be
x = yz - \uo \sqrt{1 - z^2} \ , \llabel{141230d}
\ee
with
\be
\zy = \frac{\K_0 - L \, G_2 y}{G_1 \, \gy} 
\ , \llabel{141230e0}
\ee
\be
\gy = \sqrt{(G_2 + L \, y)^2+(L \, \uo)^2}
\ . \llabel{141230e}
\ee

\subsection{Classical approximation}

\llabel{classical}

For a small satellite, we assume that $m_0 \ll m_1 < m_2$ and $R \ll a_1 \ll a_2$.
As a consequence, with $\C \sim m_0 R^2$, it follows from expressions (\ref{141216a})$-$(\ref{141216d}) that $L \ll G_1 \ll G_2$ and $\alpha_2 \ll \alpha_1$.
The Hamiltonian (\ref{141216n}) can then be simplified as
\be
\H_0 \approx \alpha_1 x^2 + \gamma z^2
\ , \llabel{141226a}
\ee
and from expression (\ref{141216o})
\be
\z \approx \z_0 - \frac{L}{G_1} \y \ , \quad \mathrm{with} \quad \z_0 = \frac{\K_0}{G_1 G_2} = cte
\ . \llabel{141226b}
\ee
Replacing $\z$ in expression (\ref{141226a}), we obtain the Hamiltonian that is often used to study Cassini states \citep[e.g.,][]{Colombo_1966, Ward_1975, Henrard_Murigande_1987}
\be
\H_0'  = \H_0 - \gamma z_0^2 
\approx \alpha_1 x^2 + 2 g y = \alpha_1 (\vs \cdot \vk_1)^2 + 2 g (\vs \cdot \vk_2)
\ , \llabel{141226c}
\ee
where 
\be
g = - z_0 \gamma L  / G_1 = cte \llabel{150104a} \ .
\ee
Here, $g/L$ is the constant precession rate of $\vk_1$ about $\vk_2$, while 
$I_0 = \cos^{-1}(\z_0) $ is the constant inclination between these two vectors.
\citet{Colombo_1966} has shown that the previous Hamiltonian represents a family of parabolas, whose interception with the unit sphere gives the possible trajectories for the spin axis, $\vs$.
In the precessing frame, one can express $\vk_1 = (0,0,1)$, $\vk_2 = (\sin I_0,0,\cos I_0)$, and $\vs = (\sin \theta \cos \phi,\sin \theta \sin \phi,\cos \theta)$, thus
\begin{eqnarray}
\H_0' &\approx& \alpha_1 \cos^2 \theta + 2 g \left(\cos I_0 \cos \theta + \sin I_0 \sin \theta \cos \phi  \right) \nonumber \\
&=& \alpha_1 (1-u^2-v^2) + 2 g \left( z_0 \sqrt{1-u^2-v^2} + \ui \sqrt{1-z_0^2} \right)
\ . \llabel{141226f}
\end{eqnarray}

If one adopts the variables ($\ui,\vi$) from the very beginning, we can obtain the same Hamiltonian without introducing the precessing frame.
Indeed, with the assumptions done for the classical approximation, we have $\zx \approx \z_0 (1 - L\x/G_1)$ (Eq.\,(\ref{141224a0})) and $\gx \approx G_1 (1 + L\x/G_1) $ (Eq.\,(\ref{141224a})). 
We then rewrite (\ref{141217f}) as
\be
z \approx z_0 \left(1-\frac{L x}{G_1} \right) - \frac{L \, u}{G_1} \sqrt{1-z_0^2} =  z_0 - \frac{L}{G_1} \left( x \, z_0 + \ui \sqrt{1-z_0^2} \right)
\ . \llabel{141226d}
\ee
Replacing $\z$ above in the general Hamiltonian (\ref{141226a}), we directly obtain expression (\ref{141226f}) again. 
The huge advantage of the new description presented in section~\ref{newvars} is that it still holds in more general situations when $L \sim G_1 \sim G_2$ and $\alpha_1 \sim \alpha_2$, for which the precession rate of $\vk_1$ about $\vk_2$ and the angle between these two vectors are no longer constant.

\section{Cassini states}

Cassini states correspond to equilibria of the spin axis.
They can thus be given by the extrema of the Hamiltonian (Eq.\,(\ref{141217g})):
\be
\frac{\partial \H_0}{\partial \ui} = 0 \quad \wedge \quad \frac{\partial \H_0}{\partial \vi} = 0 \ .
\llabel{150513a}
\ee 
Since $\H_0 = \H_0 (\x, \ui)$, we have for the derivative with respect to $\vi$
\be
\frac{\partial  \H_0}{\partial \vi} = \frac{\partial  \H_0}{\partial \x}\frac{\partial \x}{\partial \vi} = - \frac{\partial  \H_0}{\partial \x}\frac{\vi}{\x} = 0 \ .
\llabel{150514a}
\ee 
We then conclude that $\vi = 0$ is always a possible equilibrium solution (equivalent to $\phi = 0$), where the unit vectors $\vs$, $\vk_1$, and $\vk_2$ remain coplanar.
Replacing $\vi=0$ in the derivative with respect to $\ui$ (Eq.\,(\ref{150513a})) provides a general implicit condition for coplanar Cassini states:
\be
\left. \frac{\partial \H_0}{\partial \ui} \right|_{\vi=0} = - 2 \alpha_1 \uic + 2 \alpha_2 \y_c \frac{\partial y_c}{\partial \uic} + 2 \gamma \z_c \frac{\partial z_c}{\partial \uic} = 0 \ ,
\llabel{150511a}
\ee
with $\uic= \sin \theta_c $, and (Eqs.\,(\ref{141217c})$-$(\ref{141217f}))
\be
x_c = \sqrt{1-\uic^2} \ , \quad 
z_c=z(\uic, x_c) \ , \quad \mathrm{and} \quad  
y_c=y(\uic, x_c, z_c) 
\ . \llabel{150122a}
\ee
The roots of (\ref{150511a}) can be found in the interval $ \uic \in [-1,1]$ using numerical methods or simply by plotting its graph.

Alternatively, coplanar states can be obtained as stationary solutions for the equations of motion ($\dot \ui = \dot \vi = 0$), for which $\vi = 0$. 
Therefore, they can be simply obtained by setting $\vi = 0$ and $\dot \vi = 0$ (Eq.\,(\ref{141227e})) or, equivalently, for $w = 0$ and $\dot w = 0$ (Eq.\,(\ref{141216m})).
For a given value of the total angular momentum of the system, $\K_0$, the Cassini states then verify the following condition
\be
\uic = \frac{ (y_c z_c - x_c) \left(\frac{\gamma}{G_1} z_c - \frac{\alpha_2}{L} y_c \right) + (x_c y_c - z_c) \left(\frac{\alpha_2}{G_2} y_c - \frac{\alpha_1}{G_1} x_c \right) }{\sqrt{1-z_c^2} \left(\frac{\alpha_1}{L} x_c -\frac{\gamma}{G_2} z_c \right)}
\ . \llabel{141217h}
\ee

Since $\vi=\vo$ (Eq.\,(\ref{141230g})), we can obtain an equivalent condition for the coplanar Cassini states in terms of $\uoc = \cos \varepsilon_c$
\be
\left. \frac{\partial \H_0}{\partial \uo} \right|_{\vo=0} = 2 \alpha_1 x_* \frac{\partial x_*}{\partial \uoc} - 2 \alpha_2 \uoc + 2 \gamma z_* \frac{\partial z_*}{\partial \uoc} = 0 \ ,
\llabel{150618a}
\ee
or
\be
\uoc = \frac{(x_* z_* - y_*) \left(\frac{\alpha_1}{L} x_* -\frac{\gamma}{G_2} z_* \right) + (x_* y_* - z_*) \left(\frac{\alpha_2}{G_2} y_* - \frac{\alpha_1}{G_1} x_*  \right) }{\sqrt{1-z_*^2} \left(\frac{\gamma}{G_1} z_* - \frac{\alpha_2}{L} y_* \right)}
\ , \llabel{141230z}
\ee
with (Eqs.\,(\ref{141230c})$-$(\ref{141230e}))
\be
 y_* = \sqrt{1-(\uoc)^2} \ , \quad 
 z_* = z(\uoc, y_*) \ , \quad \mathrm{and} \quad
 x_* = x(\uoc, y_*, z_*)
 \ . \llabel{150122b}
\ee

\begin{table}
\caption{Observed parameters for the HAT-P-13 system \citep{Winn_etal_2010b} and the Earth-Moon system \citep{Yoder_1995cnt}. The initial conditions for the spin of HAT-P-13 are arbitrary. \llabel{Tab1}} 
\begin{center}
\begin{tabular}{|c|rl|rl|} \hline
Param. & \multicolumn{2}{|c|}{HAT-P-13} & \multicolumn{2}{|c|}{Earth-Moon} \\ \hline
$m_0$&  1.25 & \hskip-.3cm $M_\odot$& 1.00 & \hskip-.3cm $M_\oplus$  \\ 
$m_1$& 0.851 & \hskip-.3cm $\mj$& 0.0123 & \hskip-.3cm $M_\oplus$  \\ 
$m_2$& 14.28 & \hskip-.3cm $\mj$& 1.00 & \hskip-.3cm $M_\odot$ \\ 
$a_1$& 0.0427 & \hskip-.3cm a.u. & 60.34 & \hskip-.3cm $R_\oplus$  \\ 
$a_2$& 1.226 & \hskip-.3cm a.u. & 1.00 & \hskip-.3cm a.u.  \\ 
$e_1$& \multicolumn{2}{|c|}{0.0133}  & \multicolumn{2}{|c|}{0.0549}  \\ 
$e_2$& \multicolumn{2}{|c|}{0.662} & \multicolumn{2}{|c|}{0.0167}  \\ \hline
$I$ & 30. & \hskip-.3cm deg & 5.145 & \hskip-.3cm deg  \\ 
$\theta$ & 0. & \hskip-.3cm deg & 18.295 & \hskip-.3cm deg  \\ 
$\varepsilon$ & 30. & \hskip-.3cm deg & 23.44 & \hskip-.3cm deg \\ 
$P_\mathrm{rot}$ & 10. & \hskip-.3cm day & 0.997 & \hskip-.3cm day  \\ 
$R$ & 1.559 & \hskip-.3cm $R_\odot$ & 1.00 & \hskip-.3cm $R_\oplus$  \\ 
$J_2$ ($\times 10^{-6}$) & \multicolumn{2}{|c|}{3.88} & \multicolumn{2}{|c|}{1096.}  \\ 
$C / (m_0 R^2)$ & \multicolumn{2}{|c|}{0.080} & \multicolumn{2}{|c|}{0.331} \\  \hline
\end{tabular}
\end{center}
\end{table}

\subsection{Classical states}

In the classical approximation, the Hamiltonian can be simplified by $\H_0'$ (Eq.\,(\ref{141226f})), therefore
\be
\left. \frac{\partial \H_0'}{\partial \ui} \right|_{\vi=0} = - 2 \alpha_1 \uic + 2 g \left( - z_0 \frac{\uic}{\x_c} + \sqrt{1-z_0^2} \right) = 0 \ ,
\llabel{150602a}
\ee
which is equivalent to 
\be
\alpha_1 \uic \x_c + g \left( \uic z_0 - \x_c \sqrt{1-z_0^2} \right) = 0
\ . \llabel{141227b}
\ee
This expression corresponds to the commonly used condition for finding the equilibrium points for the spin axis \citep[e.g.,][]{Colombo_1966, Peale_1969, Ward_1975}.
It is usually expressed in terms of the obliquity $\uic = \sin \theta_c $, $x_c = \cos \theta_c$, and inclination $z_0 = \cos I_0$ as \citep[e.g.,][]{Ward_Hamilton_2004}
\be
\alpha_1 \sin \theta_c \cos \theta_c + g \sin(\theta_c - I_0) = 0
\ . \llabel{141227c}
\ee
This condition could also have been obtained from expression (\ref{141217h}), 
performing the same approximations as in section~\ref{classical},
\be
\uic \approx \frac{ (y_c z_c - x_c) \left(\frac{\gamma}{G_1} z_c \right)}{\sqrt{1-z_c^2} \left(\frac{\alpha_1}{L} x_c \right)} = \frac{ (y_c z_c - x_c)}{\alpha_1 x_c \sqrt{1-z_c^2}} \left(\frac{\gamma L}{G_1} z_c \right)
\ , \llabel{141227a}
\ee
since $y_c = x_c z_c + \uic \sqrt{1-z_c^2}$ (Eq.\,(\ref{141217d})), and
$z_c \approx z_0$ (Eq.\,(\ref{141226b})).

\subsection{Stars with close-in companions}

\begin{figure}[ht]
\begin{center}
\includegraphics[width=\columnwidth]{\figpath 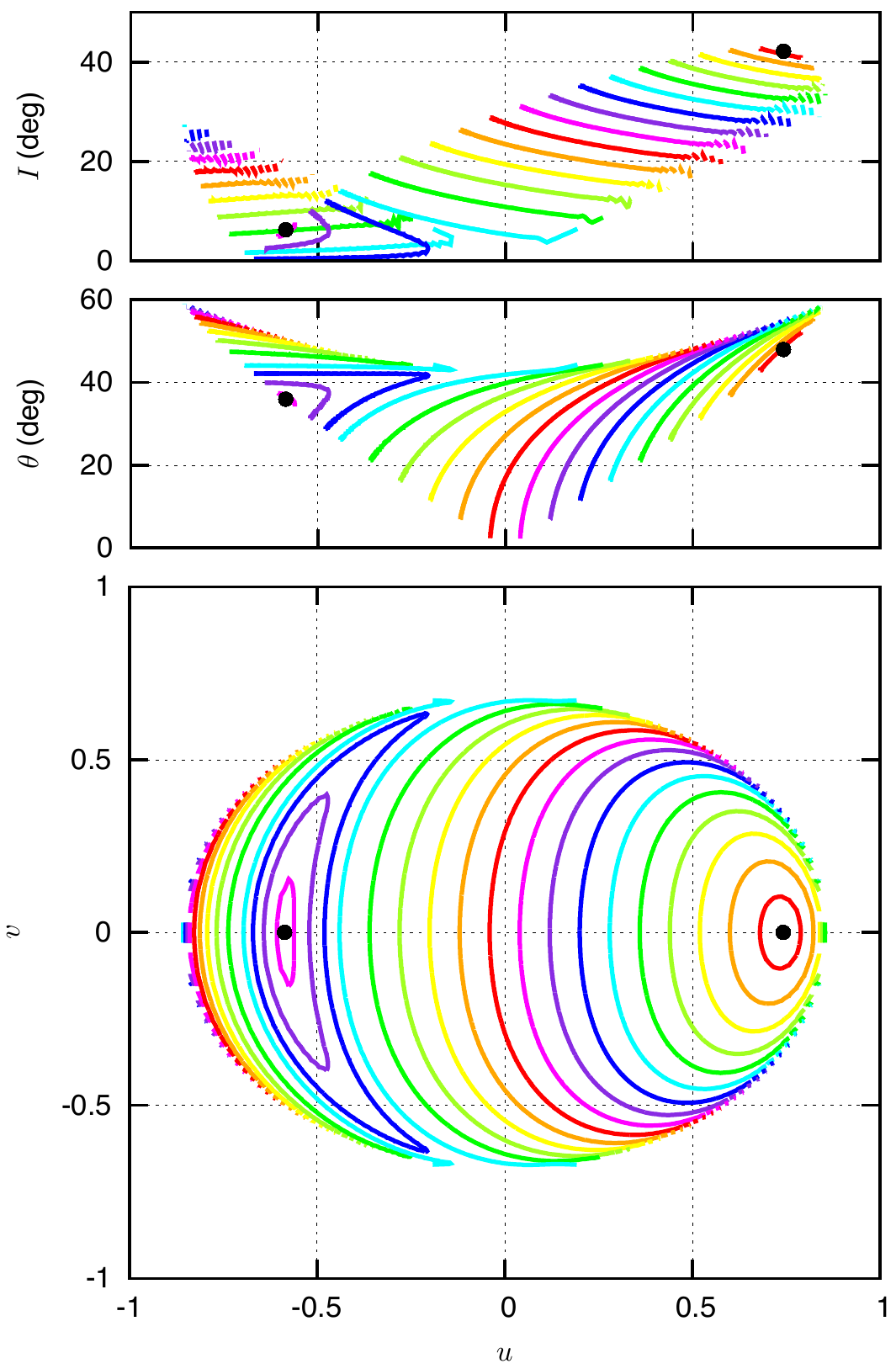} 
 \caption{Secular trajectories in the HAT-P-13 system (Table~\ref{Tab1}). We show the mutual inclination (top), the stellar spin projected on the inner orbit normal (middle), and its projection on the orbital plane (bottom). Cassini states are marked with a dot. 
 \llabel{hatp13a}  }
\end{center}
\end{figure}

\begin{figure}[ht]
\begin{center}
\includegraphics[width=\columnwidth]{\figpath 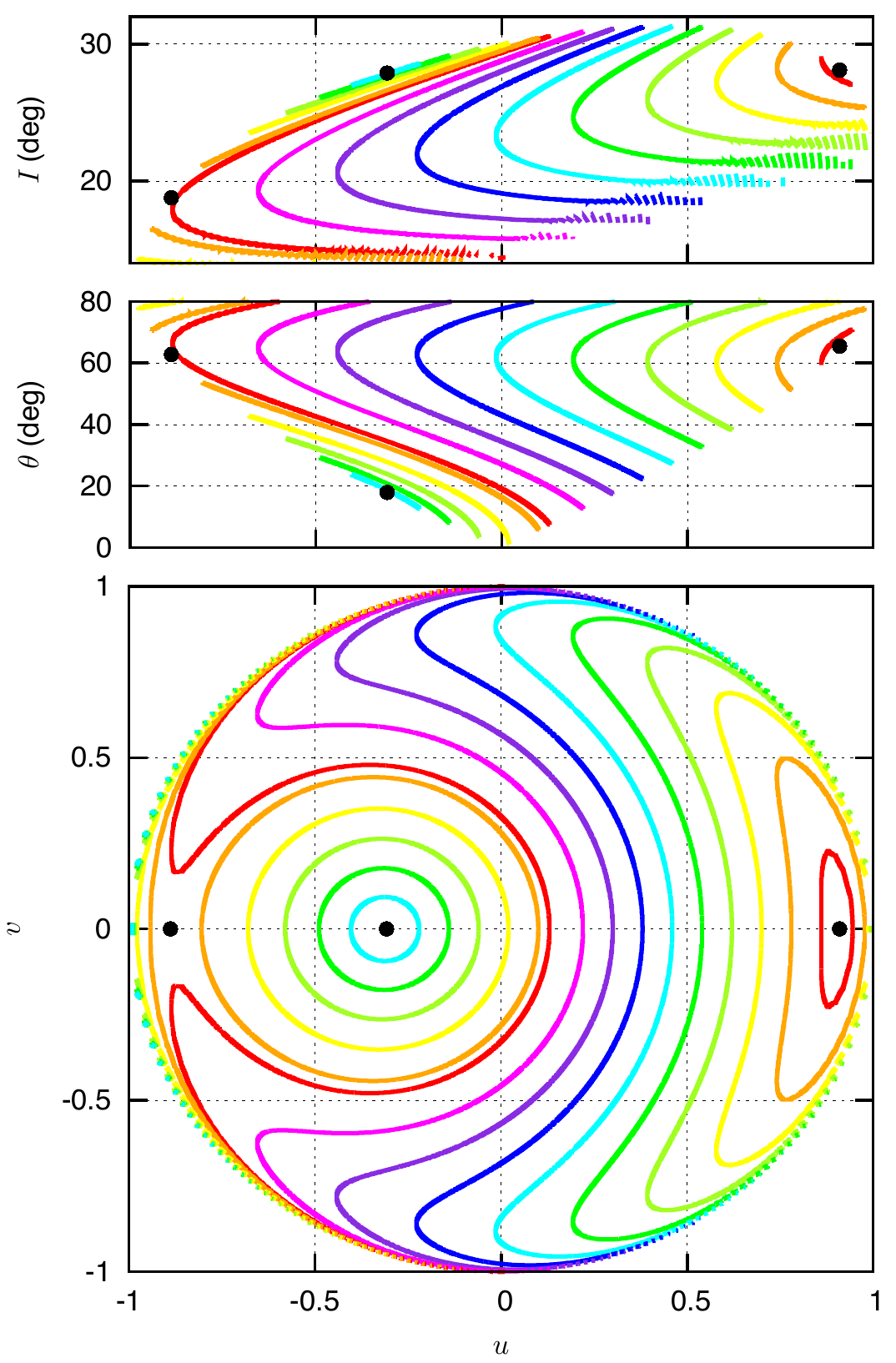} 
 \caption{Secular trajectories in the HAT-P-13 modified system with $m_1 = 8.51\,\mj$. We show the mutual inclination (top), the stellar spin projected on the inner orbit normal (middle), and on the orbital plane (bottom). Cassini states are marked with a dot. 
 \llabel{hatp13b}  }
\end{center}
\end{figure}

Unlike satellites, the rotational angular momentum of stars is often comparable to the orbital angular momentum of close-in companions.
Therefore, the classical approximation from section~\ref{classical} is not valid, and we need to apply the more general method presented in section~\ref{newvars}.

We first consider the case of a single star with two Jupiter-like planetary companions, for instance the HAT-P-13 system, for which $L \sim G_1 \lesssim G_2$ (Table~\ref{Tab1}).
This system is well constrained, since data was collected combining radial velocity and transit measurements \citep{Winn_etal_2010b}.
The inner planet is a transiting hot Jupiter in a 2.9 day quasi-circular orbit, while
the outer body has a 1.2 yr eccentric orbit and a minimum mass of about 14 Jupiter masses. 
The true mass and orbital inclination of the outer companion are unknown, so it is the mutual inclination between the two orbits.
To enhance Cassini states, we set the initial value of $I = 30^\circ$ when $\vi = 0$.
By modelling the Rossiter-McLaughlin effect, \citet{Winn_etal_2010b} also show that the inner orbit angular momentum vector and the stellar spin vector are nearly aligned on the sky, so we take the initial $\theta = 0^\circ$.
The total angular momentum of the system (Eq.\,(\ref{141216o})) is then
\begin{eqnarray}
\K_0 
& = & L \, G_1 + (L + G_1) \, G_2 \cos (30^\circ)
\ . \llabel{150115a}
\end{eqnarray}

The rotation of the star is unknown. This parameter is important for computing the rotational angular momentum of the star (Eq.\,(\ref{141216a})), but also for estimating its oblateness through
\citep[e.g.,][]{Correia_Rodriguez_2013} $ J_2 = k_2 \omega^2 R^3 / (3 \G m_0) $. 
If the rotation is synchronous with the orbital period of the inner orbit, we will get a rotation period of 2.9~days. 
However, \citet{Winn_etal_2010b} estimate the projected stellar rotation rate to be $1.66 \pm 0.37$~km/s, which gives an upper limit of 48~days for the rotation period.
Adopting $k_2 = 0.028$ \citep{Mecheri_etal_2004} and assuming a rotation period of 10~days, we have $J_2 \sim 10^{-6}$.

In Figure~\ref{hatp13a} (bottom), we show the secular trajectories for the spin projected on the inner orbit plane, obtained by plotting the level curves $H_0 (\ui,\vi,\K_0) = cte$ (Eq.\,(\ref{141217g})), that is,
without integrating the equations of motion.
We observe that there are two Cassini states, $\uic = -0.586$ and $\uic = 0.743$, which correspond to $\theta_c \approx -36^\circ$ and $\theta_c \approx 48^\circ$, respectively.
This is a striking result, because in the classical approximation, only one final state was expected.
Moreover, Cassini states correspond to the final outcome of tidal evolution, but we presently observe $\theta \approx 0^\circ$.
Therefore, either the present state is still precessing around a Cassini state (i.e., it is not yet damped), and we got it close to zero by chance, or the mutual inclination of the system is lower than $30^\circ$. 
Indeed, adopting a smaller initial inclination $I = 3^\circ$ in expression (\ref{150115a}), we get $\theta_c \approx -3.4^\circ$ and $\theta_c \approx 4.9^\circ$, which are more compatible with the observational data.

\begin{figure}
\begin{center}
\includegraphics[width=\columnwidth]{\figpath 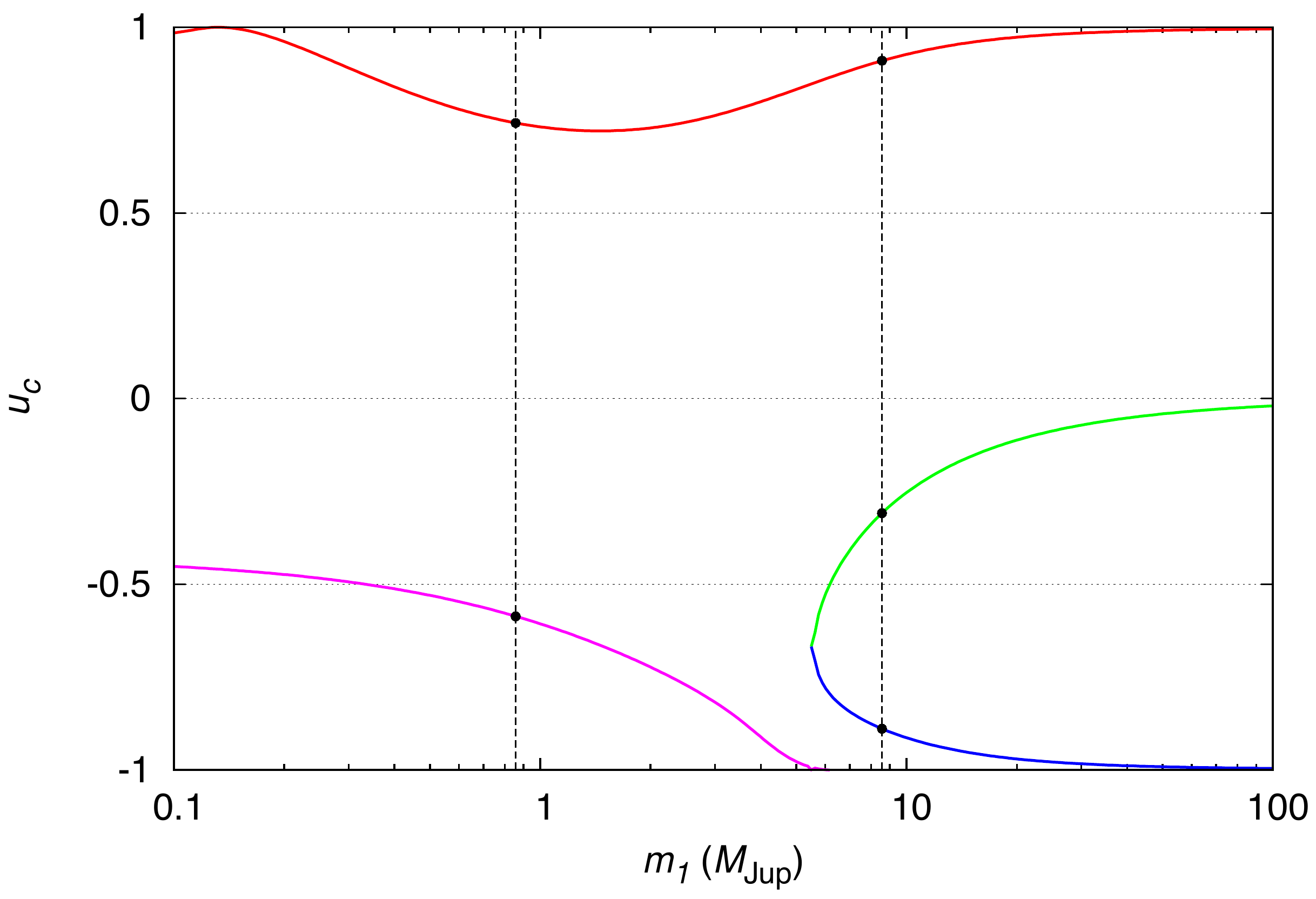} 
 \caption{Cassini states as a function of $m_1$ for a HAT-P-13-like system with initial $I = 30^\circ$. These equilibria are obtained by solving equation (\ref{150511a}). Vertical dotted lines correspond to the configurations shown in Figures~\ref{hatp13a} and \ref{hatp13b}. \llabel{cassinistates}  }
\end{center}
\end{figure}

In Figure~\ref{hatp13a} we additionally show the secular trajectories for the spin projected on the inner orbit normal $\vk_1$ (middle), obtained using $x = x(\ui,\vi)$ (Eq.\,(\ref{141217c})), and for the mutual inclination (top), obtained using $z = z(\ui,\vi,\K_0)$ (Eq.\,(\ref{141217f})).
The $(\ui,\vi)$ that we use for each trajectory are those that simultaneously verify a given level curve $H_0 (\ui,\vi,\K_0) = cte$.
Therefore, we obtain the variations in the direction cosines without integrating the equations of motion.
We observe that the mutual inclination undergoes some oscillations around the Cassini equilibria states, which was not possible in the classical approximation.
Furthermore, the two Cassini states have quite different values for $I_c \approx 6.2^\circ$ and $I_c \approx 42.2^\circ$, respectively.
Here, the conserved quantity for all trajectories is only the total angular momentum $\K_0$ and no longer $z$, contrarily to the classical description, for which $z \approx \K_0/(G_1 G_2) = cte$ (Eq.\,(\ref{141226b})).

We now consider a fictitious system with the exact same parameters as in HAT-P-13 (Table~\ref{Tab1}), but where the inner planet is ten times more massive. 
As before, we also adopt a rotation period of 10~days and initial $I = 30^\circ$, $\theta = 0^\circ$, and $\vi = 0$ (Eq.\,(\ref{150115a})).
In Figure~\ref{hatp13b} we show the secular trajectories for the modified HAT-P-13 system as in Figure~\ref{hatp13a} for the standard system.

In this case we have $L \lesssim G_1 \sim G_2$. 
The classical approximation is still inappropriate, although the level curves of the Hamiltonian recall those plotted by \citet{Ward_Hamilton_2004} for the spin of Saturn.
We count three different Cassini states, $\uic = -0.889$, $\uic = -0.308$, and $\uic = 0.910$, which correspond to $\theta_c \approx -63^\circ$, $\theta_c \approx -18^\circ$, and $\theta_c \approx 65^\circ$, respectively.
The smaller one corresponds to a hyperbolic unstable point, but the spin can be stabilised in the other two states.
However, in contrast to the classical case, here the mutual inclination undergoes significant variations.
Moreover, the Cassini states also present different values for the equilibrium inclination, $I_c \approx 18.8^\circ$, $I_c \approx 27.9^\circ$, and $I_c \approx 28.1^\circ$, respectively.

When we modify the mass of the inner planet $m_1$, we change the quantities $\alpha_1$ and $\gamma$ (Eq.\,(\ref{141216d})) and the angular momentum of the inner orbit, $G_1$ (hence $\K_0$).
Since the ratios $\alpha_1/G_1$ and $\gamma/G_1$ remain almost unchanged, different $m_1$ values only lead to different precession rates of the spin axis $\vs$ (Eq.\,(\ref{141216e})). 
In the classical approximation (section~\ref{classical}), only the variations in the precession of the spin are significant (through the term in $\alpha_1 / L$).
Therefore, modifications in the Cassini states equilibrium points are usually studied as a function of the ratio $\alpha_1/g$ (Eq.\,(\ref{141227c})).
In Figure~\ref{cassinistates} we show the Cassini states equilibria as a function of $m_1$.
Since $g$ is also constant for different $m_1$ values (Eq.\,(\ref{150104a})), this figure is equivalent to the classical maps for the ratio $\alpha_1/g  \propto m_1$ \citep[see, for instance, Fig.~3 in][]{Ward_Hamilton_2004}.
For $m_1 \gg M_\mathrm{Jup}$, the number of Cassini states is the same as in the classical case, but for lower masses, an additional Cassini state exists for $u_c < 0$, while the state for $u_c > 0$ can reach values very close to 1.

We thus see that the classical approximation from section~\ref{classical} is unable to correctly describe the secular motion of HAT-P-13-like systems. 
Some trajectories for the spin may present similarities with the classical case, but they can also present very different behaviours.
In particular, Cassini states cannot be given by expression (\ref{141227c}), since $\z$ is not constant, so we do need to find the roots when solving the more general equations (\ref{150511a}) or (\ref{141217h}).
Since Cassini states correspond to the end point of tidal evolution, if we are able to estimate the obliquity of tidally evolved stars we can put some constraints on the geometry of the orbits.

\subsection{The Earth-Moon system}

\begin{figure*}
\begin{center}
\includegraphics[width=\textwidth]{\figpath 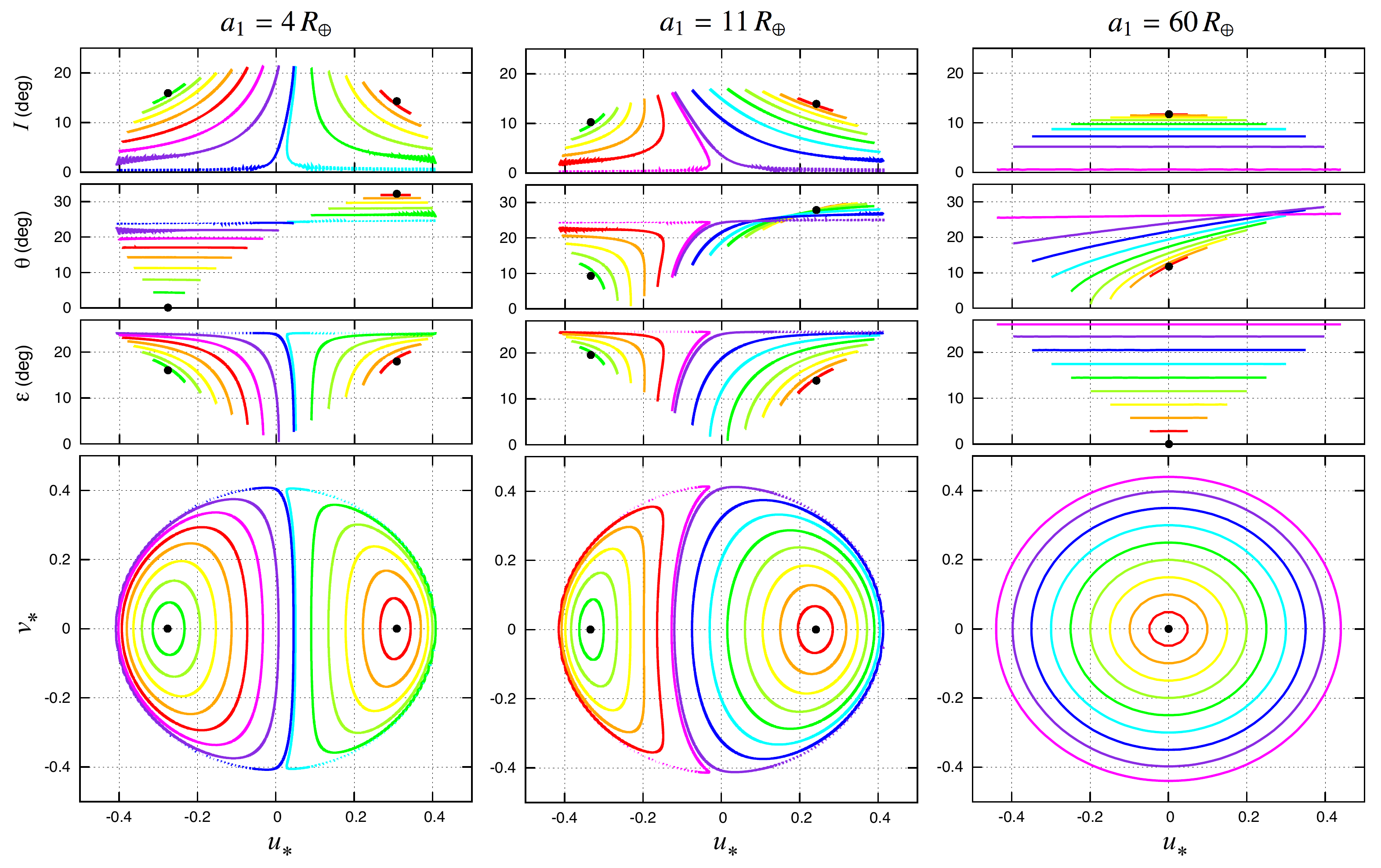} 
 \caption{Secular trajectories for the Earth-Moon system for different values of the semi-major axis, $a_1 = 4\,R_\oplus$ (left), $a_1 = 11\,R_\oplus$ (middle) and $a_1 = 60\,R_\oplus$ (right).
We show the angles between the unit vectors ($I$, $\theta$ and $\varepsilon$), together with
the projection of the Earth's spin axis on the ecliptic ($\uo,\vo$). Cassini states are marked with a dot.
To compare with Fig.~1 and 2 in \citet{Touma_Wisdom_1994} and Fig.~6 and 7 in \citet{Boue_Laskar_2006}.
 \llabel{earthmoon}  }
\end{center}
\end{figure*}

For planets possessing a massive satellite, as in the Earth-Moon system (or in the Pluto-Charon system), the rotational angular momentum of the planet can be comparable to the orbital angular momentum of the satellite. 
However, in the case of a planet, the outer orbit's angular momentum is much larger than the inner orbit's, so we have $L \sim G_1 \ll G_2$.
As a consequence, the outer orbit is almost an inertial frame, and it is more common to express the spin of the planet with respect to this plane.
We thus adopt the set of variables ($\uo, \vo$) from section~\ref{projout} here.

We first consider the present orbital configuration of the Earth-Moon system with $a_1 \approx 60\,R_\oplus$ and $I = 5.145^\circ$ (Table~\ref{Tab1}).
For the spin of the Earth we adopt the present rotation period with obliquity $\varepsilon = 23.44^\circ$ and $ J_2 \approx 10^{-3} $ \citep{Yoder_1995cnt}.
The total angular momentum of the system (Eq.\,(\ref{141216o})) is obtained setting $\vo = 0$ (which gives $\theta = 18.3^\circ$):
\be
\K_0 \approx  (L \cos (23.44^\circ) + G_1 \cos (5.145^\circ)) \, G_2
\ . \llabel{150119a}
\ee

In Figure~\ref{earthmoon} (right), we show the secular trajectories for the spin projected on the plane of the ecliptic (outer orbit).
We plot the level curves $H_0 (\uo,\vo,\K_0) = cte$ (Eq.\,(\ref{141230b})),
together with the relative positions of the unit vectors $\vs$, $\vk_1$, and $\vk_2$ (Eq.\,(\ref{141216j})).
These trajectories are obtained without integrating the equations of motion, unlike in previous studies on the Earth-Moon system. 
We observe that the spin axis of the Earth describes an almost perfect circle at constant obliquity around the Cassini state $\varepsilon_c = 22'' $, which is very close to the ecliptic pole. 
The mutual inclination between the orbit of the Moon and the ecliptic is also nearly constant, since the Laplacian plane of the Earth-Moon system almost coincides with the ecliptic \citep[e.g.,][]{Tremaine_etal_2009}.

Actually, at the present Earth-Moon distance $a_1 \approx 60\,R_\oplus$, we have $ L \ll G_1 $, so we could have used the classical approximation (section~\ref{classical}).
However, the Moon probably formed very close to the Earth: a Mars-sized body hit the nearly formed proto-Earth, blasting material into orbit around it, which accreted to form the Moon \citep[e.g.,][]{Canup_Asphaug_2001}.
In Figure~\ref{earthmoon} (left), we show the trajectories of the spin short-time after this impact, more precisely for $a_1 = 4\,R_\oplus$, which gives $L \approx G_1$ (the remaining parameters are those in Table~\ref{Tab1}).
During the early stages of the system, the Laplacian plane is close to the equatorial plane of the Earth, so the angle between the inner orbit and the equator, $\theta$, is nearly constant \citep[e.g.,][]{Goldreich_1965}.
As a consequence, the obliquity $\varepsilon$ and the mutual inclination $I$ are no longer constant, except for the Cassini states, $\uoc = -0.276$ and $\uoc = 0.308$, which correspond to $\varepsilon_c \approx -16^\circ $ and $\varepsilon_c \approx 18^\circ $, respectively.
The precession of the spin axis projected on the ecliptic is not circular, and in the second case, it does not even encircle the ecliptic pole.
In the first Cassini state, the orbit of the satellite almost coincides with the Earth's equator, $\theta_c \approx 0.08^\circ$, while in the other it keeps a significant tilt with respect to the equator, $\theta_c \approx 32^\circ$.

As a result of tidal dissipation, the Moon evolved from the primordial close-in orbit into the present one \citep[e.g.,][]{Touma_Wisdom_1994}.
On its way, there is a critical distance around $11\,R_\oplus$ where the Laplacian plane progressively shifts from the equator to the ecliptic.
In Figure~\ref{earthmoon} (middle), we show the trajectories of the spin for $a_1 = 11\,R_\oplus$.
At this evolutionary stage, none of the direction cosines (Eq.\,(\ref{141216j})) are constant unless the spin is trapped in the Cassini states, $\uoc = -0.334$ and $\uoc = 0.242$, which correspond to $\varepsilon_c \approx -20^\circ $ and $\varepsilon_c \approx 14^\circ $, respectively.
The precession of the spin axis projected on the ecliptic is similar to the previous case with $a_1 = 4\,R_\oplus$, except that the circulation area around $\varepsilon_c \approx 14^\circ $ is larger.
Indeed, as the Moon moves away from the Earth, the Cassini state with $\uoc > 0$ approaches zero, and the area around it grows, while the Cassini state with $\uoc < 0$ is shifted to the left until it disappears (Fig.\,\ref{earthmoon}, right).

As for the HAT-P-13 system from previous section, when the Moon is closer to the Earth, the rotational and the orbital angular momenta have similar magnitudes.
Therefore, the classical approximations to find the Cassini states do not work, and we need to solve the more general equations (\ref{150618a}) or (\ref{141230z}) to determine them.
Moreover, the general method presented here also provides an easy way of determining the limits for the variations in the relative positions of the unit vectors $\vs$, $\vk_1$, and $\vk_2$.
This information is very useful for climatic models when we inspect the habitability of new worlds, and it does not require performing numerical simulations.

Finally, since this method only relies on the total angular momentum of the system (Eq.\,(\ref{141216o})), it can also be used to quickly determine constraints for the past history of our planet.
For simplicity, the evolution shown in Figure~\ref{earthmoon} only accounts for variations in the semi-major axis. A more rigorous analysis requires that the conserved quantity in expression (\ref{141216o}) is $\K$ instead of $\K_0$, and that the amount of angular momentum lost in $\vG_1$ is transferred to $\vL$ by increasing the rotation rate of the Earth.

\section{Conclusions}

In this paper we have presented a simple method for determining Cassini states and the trajectories of the spin in the secular three-body problem.
This method is more general than previous approaches because it does not require the rotational angular momentum to be much smaller than the orbital one.
Therefore, it can be used to study stars with close-in companions or planets with heavy satellites, for which the precession rate and the mutual inclination of the orbits are not constant.
Our method only depends on the geometry of the Hamiltonian and thus does not require an integration of the equations of motion.

We have shown that previously unknown Cassini states may exist at high obliquities. 
As a consequence, the spin of tidally evolved stars with close-in companions can be significantly misaligned, provided that the orbits of the companions are also not coplanar.
Thus, if we are able to determine the obliquity of these stars, we can put some constraints on the relative inclination between the two orbits.
Planets with large close-in satellites can also present unexpected equilibrium configurations.

Our method relies on the conservation of the total angular momentum of the system.
Thus, it can be useful to easily track the dynamical evolution of the system when it is subject to dissipation.
Indeed, if we assume adiabatic evolution, the spin axis will travel across the constant energy levels of the Hamiltonian $H (\ui,\vi,\K_0) $ towards stable Cassini states.
We can thus predict the final configuration without performing numerical simulations.

Our model has some limitations. 
When $m_0 \sim m_1$ and $r_1 \sim R$ (for instance, a system of close binary stars), 
the spin and the $J_2$ of the companion mass $m_1$ should also be taken into account  in our analysis. 
The problem is no longer integrable, but close solutions can still be found. 
The Hamiltonian (Eq.\,(\ref{141216c})) was obtained in the frame of the quadrupolar non-restricted problem, i.e., we assumed $r_1 \ll r_2$.
Therefore, when octupole or resonant perturbations become important (close semi-major axis and/or very eccentric orbits), our method also only gives approximate results.
Finally, since we averaged the Hamiltonian over the argument of the perihelion of the inner orbit, our method is not valid when the perihelion is in libration.
This can be the case for extremely high values of the mutual inclination, where exchanges between $e_1$ and $I$ may occur \citep[e.g.,][]{Farago_Laskar_2010}.

In our model, we considered only the three-body problem.
Although this represents many situations observed in nature, planetary systems usually contain more bodies.
The n-body problem with spin is very complex, and it has a large number of degrees of freedom \citep[e.g.,][]{Boue_Fabrycky_2014a}.
However, we can generalise our method to those situations in the same way as done for the classical studies on Cassini states \citep[e.g.,][]{Ward_Hamilton_2004, Peale_2006}.
The secular perturbations of a n-body system on the inner orbit can be decomposed in quasi-periodic series of the secular forcing frequencies in the system \citep[e.g.,][]{Laskar_1988}.
For an isolated term associated with a specific frequency, the Hamiltonian can be simplified and made integrable as shown here.
Cassini states can therefore be found for the dominating perturbations in the Hamiltonian.

\begin{acknowledgements}
A.C. thanks G. Bou\'e and J. Laskar for discussions, 
and acknowledges support from CIDMA strategic project UID/MAT/04106/2013.
\end{acknowledgements}

\bibliographystyle{aa}
\bibliography{correia}

\begin{thebibliography}{23}
\expandafter\ifx\csname natexlab\endcsname\relax\def\natexlab#1{#1}\fi

\bibitem[{{Bou{\'e}} \& {Fabrycky}(2014)}]{Boue_Fabrycky_2014a}
{Bou{\'e}}, G. \& {Fabrycky}, D.~C. 2014, \apj, 789, 110

\bibitem[{{Bou{\'e}} \& {Laskar}(2006)}]{Boue_Laskar_2006}
{Bou{\'e}}, G. \& {Laskar}, J. 2006, Icarus, 185, 312

\bibitem[{{Breiter} {et~al.}(2005){Breiter}, {Nesvorn{\'y}}, \&
  {Vokrouhlick{\'y}}}]{Breiter_etal_2005b}
{Breiter}, S., {Nesvorn{\'y}}, D., \& {Vokrouhlick{\'y}}, D. 2005, \aj, 130,
  1267

\bibitem[{{Canup} \& {Asphaug}(2001)}]{Canup_Asphaug_2001}
{Canup}, R.~M. \& {Asphaug}, E. 2001, \nat, 412, 708

\bibitem[{{Cassini}(1693)}]{Cassini_1693}
{Cassini}, G.~D. 1693, {Trait{\'e} de l'origine et du progr{\`e}s de
  l'astronomie} (Paris: Gauthier-Villars)

\bibitem[{{Colombo}(1966)}]{Colombo_1966}
{Colombo}, G. 1966, \aj, 71, 891

\bibitem[{{Correia} \& {Laskar}(2003)}]{Correia_Laskar_2003I}
{Correia}, A.~C.~M. \& {Laskar}, J. 2003, Icarus, 163, 24

\bibitem[{{Correia} \& {Rodr{\'{\i}}guez}(2013)}]{Correia_Rodriguez_2013}
{Correia}, A.~C.~M. \& {Rodr{\'{\i}}guez}, A. 2013, \apj, 767, 128

\bibitem[{{Dullin}(2004)}]{Dullin_2004}
{Dullin}, H.~R. 2004, Reg. Chaot. Dynam., 9, 255?264

\bibitem[{{Farago} \& {Laskar}(2010)}]{Farago_Laskar_2010}
{Farago}, F. \& {Laskar}, J. 2010, \mnras, 401, 1189

\bibitem[{{Goldreich}(1965)}]{Goldreich_1965}
{Goldreich}, P. 1965, \aj, 70, 5

\bibitem[{{Goldreich}(1966)}]{Goldreich_1966a}
{Goldreich}, P. 1966, Reviews of Geophysics and Space Physics, 4, 411

\bibitem[{{Henrard} \& {Murigande}(1987)}]{Henrard_Murigande_1987}
{Henrard}, J. \& {Murigande}, C. 1987, Celestial Mechanics, 40, 345

\bibitem[{{Laskar}(1988)}]{Laskar_1988}
{Laskar}, J. 1988, \aap, 198, 341

\bibitem[{{Mecheri} {et~al.}(2004){Mecheri}, {Abdelatif}, {Irbah}, {Provost},
  \& {Berthomieu}}]{Mecheri_etal_2004}
{Mecheri}, R., {Abdelatif}, T., {Irbah}, A., {Provost}, J., \& {Berthomieu}, G.
  2004, \solphys, 222, 191

\bibitem[{{Peale}(1969)}]{Peale_1969}
{Peale}, S.~J. 1969, \aj, 74, 483

\bibitem[{{Peale}(2006)}]{Peale_2006}
{Peale}, S.~J. 2006, Icarus, 181, 338

\bibitem[{{Touma} \& {Wisdom}(1994)}]{Touma_Wisdom_1994}
{Touma}, J. \& {Wisdom}, J. 1994, \aj, 108, 1943

\bibitem[{{Tremaine} {et~al.}(2009){Tremaine}, {Touma}, \&
  {Namouni}}]{Tremaine_etal_2009}
{Tremaine}, S., {Touma}, J., \& {Namouni}, F. 2009, \aj, 137, 3706

\bibitem[{{Ward}(1975)}]{Ward_1975}
{Ward}, W.~R. 1975, \aj, 80, 64

\bibitem[{{Ward} \& {Hamilton}(2004)}]{Ward_Hamilton_2004}
{Ward}, W.~R. \& {Hamilton}, D.~P. 2004, \aj, 128, 2501

\bibitem[{{Winn} {et~al.}(2010){Winn}, {Johnson}, {Howard}, {Marcy}, {Bakos},
  {Hartman}, {Torres}, {Albrecht}, \& {Narita}}]{Winn_etal_2010b}
{Winn}, J.~N., {Johnson}, J.~A., {Howard}, A.~W., {et~al.} 2010, \apj, 718, 575

\bibitem[{{Yoder}(1995)}]{Yoder_1995cnt}
{Yoder}, C.~F. 1995, in Global Earth Physics: A Handbook of Physical Constants
  (American Geophysical Union, Washington D.C), 1--31

\end{thebibliography}

\end{document}